\documentclass[prd,aps,twocolumn,preprintnumbers,nofootinbib,showpacs,amsmath,amssymb,superscriptaddress]{revtex4-1}
\usepackage{color}
\definecolor{purple}{rgb}{0.7,0.0,0.5}
\usepackage[bookmarks = false, colorlinks = true, linkcolor = blue, citecolor = purple, urlcolor=purple]{hyperref}

\usepackage{amsfonts}
\usepackage{amsmath}
\usepackage{amssymb}
\usepackage{upgreek}
\usepackage{bm}
\usepackage{dcolumn}
\usepackage{epsfig}
\usepackage{graphicx}
\usepackage{graphics}
\usepackage[latin1]{inputenc}
\usepackage{latexsym}
\usepackage{rotating}
\usepackage{hyperref}
\usepackage[all]{hypcap}
\usepackage{xspace}

\makeatletter
\AtBeginDocument{\@ifpackageloaded{natbib}{\ifNAT@numbers\if@filesw\immediate\write\@auxout{\string\global\string\NAT@numberstrue}\fi\fi}{}}
\makeatother

\begin{document}
\title{
Quantum radiation produced by a uniformly accelerating charged particle
in thermal random motion
}
\author{
Naritaka Oshita}
\affiliation{
Department~of~Physics, Graduate~School of Science, The University of Tokyo, Bunkyo-ku, Tokyo~113-0033, Japan, and Research Center for the Early Universe (RESCEU), Graduate School of Science, The University of Tokyo, Bunkyo-ku, Tokyo 113-0033, Japan}

\author{
Kazuhiro Yamamoto}
\affiliation{
Department of Physical Science, Graduate School of Science, Hiroshima University,
         Higashi-Hiroshima 739-8526, Japan, and 
Hiroshima Astrophysical Science Center, Hiroshima University,
         Higashi-Hiroshima 739-8526, Japan}

\author{Sen Zhang}
\affiliation{
Okayama~Institute~for~Quantum~Physics,
Kyoyama~1-9-1,~Kita-ku,~Okayama~700-0015,~Japan
}

\begin{abstract}
We investigate the properties of quantum radiation produced by a uniformly accelerating charged 
particle undergoing thermal random motions,  which originates from the coupling to the 
vacuum fluctuations of the electromagnetic field. The thermal random motions are 
regarded to result from the Unruh effect, the quantum radiation might give us hints of the 
Unruh effect.
The energy flux of the quantum radiation is negative and smaller than that of Larmor radiation by one order 
in $a/m$,  where $a$ is the constant acceleration  and $m$ is the mass of the particle. 
Thus, the quantum radiation appears to be a suppression of the classical Larmor radiation. 
The quantum interference effect plays an important role in this unique signature. 
The results is consistent with the predictions of a model consisting of a particle 
coupled to a massless scalar field as well as those of the previous studies on the 
quantum effect on the Larmor radiation. 
\end{abstract}
\maketitle
\def\Omega{\sigma}

\section{Introduction}
One of the most exciting phenomena related to quantum fields in non-inertial frame is the Unruh effect.
The Unruh effect is the theoretical prediction that an accelerating observer sees the Minkowski 
vacuum as a thermally excited state with the Unruh temperature $T_U={a/2\pi}$ as the natural unit, 
where $a$ is the acceleration~\cite{Unruh} (see \cite{Higuchi} for a review). 
Through the principle of equivalence, the Unruh effect is related to the Hawking effect which 
predicts radiation with a thermal spectrum from a black hole~\cite{HawkingRadiation}. 
The uniformly accelerated observer of the Unruh effect is an approximation to an observer at 
a fixed distance near the horizon of a black hole. In both cases, the observer perceives a horizon.

Although direct experimental verification of the Hawking effect seems to be difficult, that of 
the Unruh effect might be possible.  Chen and Tajima proposed a possible experimental test of 
the Unruh effect using an intense laser field ~\cite{ChenTajima}, which has inspired many studies, 
such as  Ref.~\cite{ELI,Schutzhold,Schutzhold2}. These studies suggested that the Unruh effect may give rise 
to quantum radiation from an accelerating charged particle, which is termed Unruh radiation. 
However, the problem is not entirely 
straightforward; it has been argued that the naively expected quantum radiation produced by the detector 
models cancels out due to the interference effect~\cite{Raine,Raval,IYZ13}. On the other hand, 
the quantum radiation from a uniformly accelerating charged particle may exist because the cancellation 
is partial: the naively expected radiation term cancels out but the interference terms remain~\cite{IYZ}.

Recently, we have re-investigated the quantum radiation produced by a uniformly accelerating charged 
particle coupled to vacuum fluctuations~\cite{OYZ15}. In this previous study, we adopted a model 
consisting of a particle and a massless scalar field, from which we verified that the remaining 
interference terms may give rise to a unique signature of the Unruh effect contained in the radiation. 
In the present paper, we extend our previous work to a realistic model consisting of a charged particle 
and an electromagnetic field with vector-type coupling. We demonstrate that the previously mentioned 
partial cancellation still occurs, and the remaining interference terms indeed give rise to a 
unique signature of the Unruh effect contained in the energy flux. 

It is useful to clarify the feature of our work and the difference 
between our approach and those of other previous works. 
First, our model consists of a charged particle and an electromagnetic field. 
This is the same as the previous works \cite{ChenTajima,Schutzhold,Schutzhold2,IYZ}, 
which investigated the Unruh radiation. However, the completely different point 
of our work compared with the previous works \cite{ChenTajima,Schutzhold,Schutzhold2} 
is that we take into account the interference term between the quantum vacuum 
fluctuations of electromagnetic field, $A^\mu_{\rm h}(x)$, 
and the component of electromagnetic field generated by the thermal random 
motions of a charged particle due to the Unruh effect, $A^\mu_{\rm inh}(x)$ 
(see Eq.~(\ref{solutiona}) for the definition). 
They are obtained by solving the first principle equations of motion. 
The second point is that we compute the expectation value of the energy 
momentum tensor of the electromagnetic field. 
Thus, our approach is based on a straightforward method for clear interpretation of results avoiding ambiguity.  

\section{Model}
We consider the theoretical model described in Ref.~\cite{IYZ}, which consists of 
a charged particle and an electromagnetic field. The authors of Ref.~\cite{IYZ} have shown that 
the energy equipartition relation with the Unruh temperature $T_U=a/2\pi$ appears in the random 
motions of an accelerated charged particle due to the coupling to the electromagnetic field, 
similar to the case with a massless scalar field. We focus our investigation on the quantum 
radiation from  the charged particle. The action of the model is given by 
\begin{eqnarray}
S=S_{\rm P}(z)+S_{\rm EM}(A)+S_{\rm int}(z,A),
\end{eqnarray}
where $S_{\rm P}(z)$ and $S_{\rm EM}(A)$ are the actions of the free particle and the vector field, 
\begin{eqnarray}
&&S_{\rm P}(z)=-m\int d\tau \sqrt{\eta_{\mu\nu} {\dot z}^\mu \dot z^\nu}
\label{1-1} 
\\
 &&S_{\rm EM}(A) =-{1\over4}\int {d^4x}  F^{\mu\nu}  F_{\mu\nu},
\label{1-2}
\end{eqnarray}
respectively, and $S_{\rm int}(z,A)$ describes the interaction, 
\begin{eqnarray}
 S_{\rm int}(z,A) 
  =  -e\int d\tau \int{d^4x} \delta^{4}_D\left(x-z(\tau)\right) \dot z^\mu(\tau)A_\mu(x),
\nonumber 
\\
\label{1-3}
\end{eqnarray}
where $e$ is the charge of the particle and $F_{\mu\nu}(=\partial_\mu A_\nu-\partial_\nu A_\mu)$
is the field strength.  We follow the metric convention $(+---)$.  
The equations of motion are 
\begin{eqnarray}
&&m\ddot z_\mu=e(\partial_\mu A_\nu-\partial_\nu A_\mu)\dot z^\nu+f_\mu,
\label{equationp}
\\
&&\partial_\mu \partial^\mu A^{\nu}=e\int d\tau \dot z^\nu(\tau)\delta^4_D(x-z(\tau)),
\label{equationf}
\end{eqnarray}
where we adopted the gauge condition, $\partial_\mu A^\mu=0$ and introduced an external force $f_\mu$ for a uniform acceleration. The solution to Eq.~(\ref{equationf}) is given by a combination of the homogeneous solution $A^\mu_{{\rm h}}(x)$, which satisfies $\partial_\nu \partial^\nu A^\mu_{{\rm h}}(x)=0$,  and the inhomogeneous solution written with the retarded Green's function $G_R(x,y)$, which satisfies $\partial_\nu \partial^\nu G_R(x,y)=\delta^4_D(x-y)$, as
\begin{eqnarray}
A^\mu(x)
&=&A^\mu_{{\rm h}}(x)+e\int d\tau G_R(x,z(\tau))\dot z^\mu(\tau)
\nonumber
\\
&=&A^\mu_{{\rm h}}(x)+A^\mu_{{\rm inh}}(x).
\label{solutiona}
\end{eqnarray}

Inserting the above solution into the equation of motion of the particle (\ref{equationp}), the homogeneous solution gives rise to the random force, while
the inhomogeneous solution leads to the Abraham-Lorentz-Dirac radiation reaction force, and we also have the stochastic equation of motion (see Eq.~(5.9) in Ref.~\cite{IYZ}). 
To consider random motions around uniformly accelerated motion,  we write 
\begin{eqnarray}
z^\mu(\tau)=\bar z^\mu(\tau)+\delta z^\mu(\tau),
\end{eqnarray}
where $\bar z^\mu(\tau)$ describes the uniformly accelerated motion and $\delta z^\mu(\tau)$ denotes the small perturbed random motions. The uniformly accelerated motion yields a hyperbolic trajectory written as
\begin{eqnarray}
&&\bar z^\mu(\tau)=a^{-1}(\sinh a\tau,\cosh a\tau,0,0).
\end{eqnarray}
Unless otherwise noted, we adopt the convention that the Greek letters run from 0 to 3, the Latin letters take on the values $2$ and $3$, the components of the transverse direction, and the capital Latin letters take on $0$ and $1$, the components of the longitudinal direction.  It is useful to note that $\dot {\bar z}^A$ and ${\bar z}^A$ are related to each other as
\begin{eqnarray}
 \dot {\bar z}^A= a\epsilon^A{}_{B}\bar z^B,
\end{eqnarray}
where $\epsilon_{AB}$ is the two-dimensional Levi-Civita completely antisymmetric tensor, which is defined by $\epsilon_{01}=-\epsilon_{10}=1$. 

We solve the stochastic equation using the perturbative method by expanding it with respect to $\delta z^\mu$. 
Because the random motions in the transverse direction satisfy the energy equipartition relation~\cite{IYZ}, 
we assume that the quantum radiation from the transverse fluctuations are related to the Unruh effect. 
Therefore, we restrict our investigation to the transverse motion with the $\delta z^i$ with $i=2$ and $3$. 
Expanding the stochastic equation perturbatively to first order yields 
 \begin{eqnarray}
&&m\ddot  {\delta z}_i(\tau)={e^2\over 6\pi}(\dddot {\delta z}_i-a^2\dot {\delta z}_i)
\nonumber\\
&&~~~~~~
+e(\eta_{i\nu}\dot{\bar z}_\alpha-\eta_{i\alpha}\dot{\bar z}_\nu)\partial^\nu A_{\rm h}^\alpha(x)
\big|_{x=z(\tau)}.
\end{eqnarray}

In the present paper, for simplicity, we drop the third-order time derivative term of the radiation 
reaction force. As discussed in Ref.~\cite{OYZ15}, the contribution of this term to the solution of 
$\delta z^i$ is small, and is limited to the order of ${\cal O}\bigl((a/m)^2\bigr)$.  
In that study, it is also shown that the contribution comes from the short-distance dynamics about 
the classical electron radius, $r_e= e^2/m$, which is much smaller than the Compton length. 
Assumption of the point particle is no longer valid when describing such short-distance behavior, 
where one needs to employ a more sophisticated model on the basis of wave packet~\cite{Zhang:2013ria}.  
Hence, we ignore such a term in our description of the point particle.

Then, the equation is solved by introducing the Fourier transform
\begin{eqnarray}
\delta \dot z^i(\tau)=\int {d\omega\over 2\pi} \widetilde {\delta \dot z}^i(\omega) 
e^{-i\omega\tau}, 
\end{eqnarray}
which leads to the solution
\begin{eqnarray}
 \widetilde {\delta \dot z }^i(\omega) =eh(\omega)\widetilde{
\partial^i_\alpha A^\alpha}(\omega)
\end{eqnarray}
with
\begin{eqnarray}
 h(\omega)
={1\over m(a\sigma-i\omega)} 
\end{eqnarray}
and $\sigma=e^2a/6\pi m$, and 
\begin{eqnarray}
\widetilde{\partial^i_\alpha A^\alpha}(\omega)
=\int {d\tau} e^{i\omega\tau}(\dot {\bar z}_\alpha\eta^{i\nu}-\eta^i{}_\alpha\dot{\bar z}^\nu)
\partial_\nu A_{\rm h}^\alpha(x)\Big|_{x=\bar z(\tau)}.
\end{eqnarray}
One can demonstrate that the solution satisfies the energy equipartition relation with the Unruh temperature $T_U=a/2\pi$~\cite{IYZ}
\begin{eqnarray}
{m\over 2}\bigl\langle \delta \dot z^i(\tau)\delta \dot z^j(\tau)\bigr\rangle=
{\delta_{ij}\over 2}{a\over 2\pi} \left(1+{\cal O}\left({a^2\over m^2}\right)\right)
\end{eqnarray}
by using the Wightman function
\begin{equation}
\big\langle A_{\rm h}^\alpha(x)A_{\rm h}^\beta(y)\big\rangle ={1\over 4\pi^2} {\eta^{\alpha\beta} \over
(x^0-y^0-i\epsilon)^2-({\bf x}-{\bf y})^2}.
\end{equation}

\begin{figure}[t]
 \begin{minipage}{0.5\hsize}
  \begin{center}
   \includegraphics[width=80mm,bb=80 0 600 480]{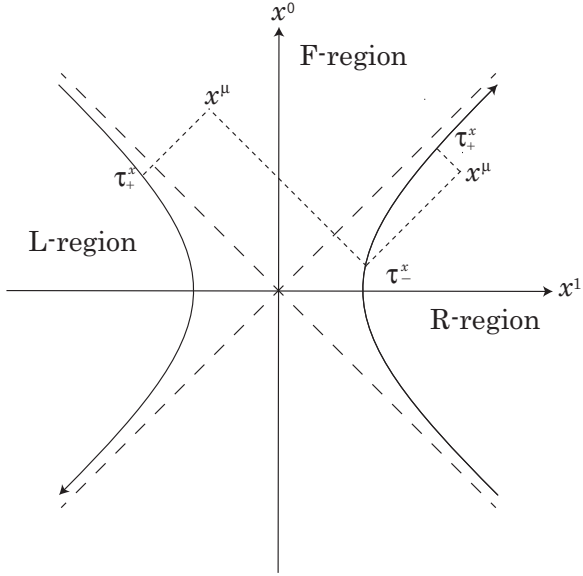}  
\end{center}
\vspace{-0.5cm}
  \label{fig:two}
 \end{minipage}
 \caption{
The hyperbolic curve in the R-region is the trajectory of a uniformly accelerating particle. The hyperbolic curve in the L-region is the hypothetical trajectory obtained by an analytic continuation of the true trajectory. For an observer at point $x^\mu$ in the R-region, $\tau_-^x$ is defined by the proper time of the particle's trajectory intersecting with the past light cone, while $\tau_+^x$ is similarly defined with the future light cone when $x^\mu$ is in the R-region. For an observer in the F-region, $\tau_+^x$ is the proper time of the hypothetical trajectory in the L-region intersecting with the past light cone.}
\end{figure}

\section{Two-Point function}
We now consider the radiation produced by the charged particle. 
To evaluate the expectation value of the energy-momentum tensor, 
we first derive the two-point function of the vector field:
\begin{eqnarray}
&&
\langle A^\alpha(x)A^\beta(y)\rangle-
\langle A_{\rm h}^\alpha(x)A_{\rm h}^\beta(y)\rangle
=\langle A_{\rm h}^\alpha(x)A_{\rm inh}^\beta(y)\rangle
\nonumber\\
&&~~~~~~
+\langle A_{\rm inh}^\alpha(x)A_{\rm h}^\beta(y)\rangle
+\langle A_{\rm inh}^\alpha(x)A_{\rm inh}^\beta(y)\rangle.
\end{eqnarray}
Using the expression for the retarded Green's function, 
$G_R(x-y)=\delta_D((x-y)^2)\theta(x^0-y^0)/2\pi$,
the inhomogeneous solution of $A^\mu(x)$ reduces to 
\begin{eqnarray}A_{\rm inh}^\mu(x)=
{e\dot z^\mu(\tau_-^x)\over 4\pi\rho(\tau_-^x)},
\end{eqnarray}
where we defined
$\rho(x)=\dot z_\mu(\tau_-^x)(x^\mu-z^\mu(\tau_-^x))$, 
and $\tau_-^x$ satisfies $(x-z(\tau_-^x))^2=0$. 
Following the perturbative expansion, $z^\mu=\bar z^\mu+\delta z^\mu$, 
we may write 
\begin{eqnarray}
\rho(x)=\rho_0(x)+\delta\rho(x)
\end{eqnarray}
with 
\begin{eqnarray}
&&\rho_0(x)=\dot {\bar z}_\mu(\tau_-^x)x^\mu,
\\
&&\delta \rho(\tau_-^x)=\delta \dot z_\mu(\tau_-^x)(x^\mu-\bar z^\mu(\tau_-^x)), 
\end{eqnarray}
where $\tau_-^x$ is redefined to satisfy $(x-\bar z(\tau_-^x))^2=0$.
Then, the inhomogeneous solution is written up to the first-order perturbative term as follows, 
\begin{eqnarray}
A^\mu_{\rm inh}(x)
={e\over 4\pi \rho_0(x)}\bigl(\dot{\bar z}^\mu(\tau_-^x)-E_{(-)}^{\mu}{}_i(x)\delta \dot z^i(\tau_-^x)\bigr),
\end{eqnarray}
which is obtained from $\delta \rho(\tau_-^x)=\delta \dot z_i(\tau_-^x)x^i$
and $\delta \dot z _\mu(\tau_-^x)=\eta_{\mu}{}^i\delta \dot z_i(\tau_-^x)$ 
for the transverse fluctuations, where we introduced
\begin{eqnarray}
E^{\mu i}_{(\mp)}{}(x)=\eta^{\mu}{}^i-{\dot{\bar z}^\mu(\tau_\mp^x)x^i \over \rho_0(x)}.
\end{eqnarray} 
Here, we also introduced $\tau_+^x$, which satisfies $(x-\bar z(\tau_+^x))^2=0$. 
The meaning of $\tau_-^x$ and $\tau_+^x$ is explained in Figure 1. 

It is straightforward to evaluate the two-point function; the explicit expression for the symmetrized two-point function with respect to $x$ and $y$ is 
\begin{eqnarray}
&&[\big\langle A^\alpha(x)A^\beta(y) - A_{\rm h}^\alpha(x)A_{\rm h}^\beta(y)
\big\rangle]_S 
=
\biggl({e\over 4\pi}\biggr)^2
{\dot{\bar z}^\alpha(\tau_-^x)\over \rho_0(x)}
{\dot{\bar z}^\beta(\tau_-^y)\over \rho_0(y)}
\nonumber\\
&&~+\biggl[{e\over 4\pi \rho_0(y)}
{e\over 4\pi\rho_0(x)}
{E_{(-)}^{\beta}{}_i(y)E_{(+)}^{\alpha i}(x)
\over 2m}I_2(x,y)
\nonumber
\\
&&~+{e\over 4\pi \rho_0(y)}
{e\over 4\pi\rho_0^3(x)}
E_{(-)}^\beta{}_{i}(y)
x^i\eta^{\alpha A}a\epsilon_A{}^{A'}x_{A'}
\nonumber\\
&&~{i\over 2m}(I_1(x,y)-I_3(x,y))\biggr]
+\biggl[(x,\alpha)\leftrightarrow(y,\beta)\biggr],
\label{twopointfunction}
\end{eqnarray}
where $I_\ell$ with $\ell=1\sim3$ is the same as Eqs.~(4.45)$\sim$(4.47) in Ref.~\cite{OYZ15} but with $\sigma=e^2a/6\pi m$.  The approximate expression for $I_\ell$ is given as (see Ref.~\cite{OYZ15}),  
\begin{eqnarray}
&&I_1(x,y)=
-{i\over 2\pi\sigma}+{i\over \pi}\log (1+e^{-a|\tau_-^y-\tau_+^x|})
\nonumber
\\
&&~~~~~~~~~~+{i\over \pi}a(\tau_-^y-\tau_+^x)\theta(\tau_-^y-\tau_+^x)+{\cal O}(\sigma),
\\
&&I_2(x,y)=-{a\over \pi} {1\over e^{a(\tau_+^x-\tau_-^y)}+1}+{\cal O}(\sigma),
\\
&&I_3(x,y)=
-{i\over 2\pi\sigma}+{i\over \pi}\log (1-e^{-a|\tau_-^y-\tau_-^x|})
\nonumber
\\
&&~~~~~~~~~~+{i\over \pi}a(\tau_-^y-\tau_-^x)\theta(\tau_-^y-\tau_-^x)+{\cal O}(\sigma),
\end{eqnarray}
for the F-region $x^0>|x^1|$ (see Fig.~1).

Here we note some details in deriving Eq.~(\ref{twopointfunction}). 
In evaluating 
$\langle A_{\rm h}^\alpha(x)A_{\rm inh}^\beta(y)\rangle
+\langle A_{\rm inh}^\alpha(x)A_{\rm h}^\beta(y)\rangle
+\langle A_{\rm inh}^\alpha(x)A_{\rm inh}^\beta(y)\rangle$, we find that 
the term $\langle A_{\rm inh}^\alpha(x)A_{\rm inh}^\beta(y)\rangle$ completely 
cancels out. Therefore, Eq.~(\ref{twopointfunction}) comes from the remaining 
interference term of $\langle A_{\rm h}^\alpha(x)A_{\rm inh}^\beta(y)\rangle
+\langle A_{\rm inh}^\alpha(x)A_{\rm h}^\beta(y)\rangle$. 
Thus, the interference term screens the radiation field carried by 
$\langle A_{\text{inh}}^\alpha(x) A_{\text{inh}}^\beta(y) \rangle$. 
This means that the component of electromagnetic field generated by the thermal 
random motions of a charged particle due to the Unruh effect cancels by the 
interference term. 
This feature is in common with the model consisting of a particle and a 
scalar field \cite{OYZ15} as well as the model consisting 
of a detector and a scalar field \cite{LinHu}.

\begin{figure}[t]
\begin{center}
 \hspace{-1cm}
  \includegraphics[width=70mm]{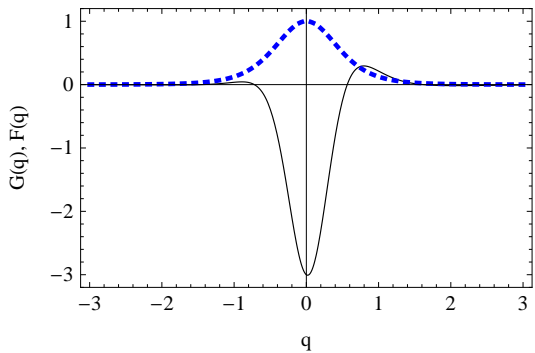}
\caption{Functions $G(q)$ (dashed curve) and $F(q)$ (solid curve) plotted without  
the divergent terms. }
  \end{center}
\end{figure}

\def\bfxperp{{x_ix^i}}
\section{Flux}
The energy flux is given by the time-space component of the energy-momentum tensor, 
\begin{eqnarray}
T_{0\mu}
=-(A_{\alpha}{}_{,0}-A_{0}{}_{,\alpha})(A^{\alpha}{}_{,\mu}-A_{\mu}{}^{,\alpha}),
\label{T0i}
\end{eqnarray}
whose expectation value is derived by differentiating the two-point function
and taking the coincidence limit, i.e.,
\begin{eqnarray}
\lim_{y\rightarrow x}
{\partial\over \partial x^\mu}{\partial\over \partial y^\nu}
[\big\langle A^\alpha(x)A^\beta(y)
-A_{\rm h}^\alpha(x)A_{\rm h}^\beta(y)\big\rangle]_S. 
\end{eqnarray}
The energy flux at large distances is obtained from the energy-momentum tensor by 
$-\sum_{i=1}^3T_{0i}n^i$ with $n^i=x^i/r$. We consider here the energy flux in the F-region, which is relevant at $r\rightarrow\infty$. The energy flux is a combination of the classical part $f^C$ and the quantum part $f^Q$, which are given by
\begin{eqnarray}
&&f^C=
\biggl({e\over 4\pi}\biggr)^2{a^2}
{1\over r^2}{1\over \sin^4\theta}G(q),
\\
&&f^Q
=
\biggl({e\over 4\pi}\biggr)^2{a^3\over 2\pi m}
{1\over r^2}{1\over \sin^4\theta}F(q),
\label{fluxc}
\end{eqnarray}
with 
\begin{eqnarray}
&&G(q)={1-P^2\over (1+q^2)^2},
\\
&&F(q)={1\over (1+q^2)^3}
\biggl[
6P(2P^2-1)
\biggl\{\log a \varepsilon
-\log (1
\nonumber\\
&&~~
+e^{-a|\tau_--\tau_+|})
-a(\tau_--\tau_+)\theta(\tau_--\tau_+)
\biggr\}
+{2P\over (a\varepsilon)^2}
\nonumber\\
&&~~
+2{(3-e^{a(\tau_+-\tau_-)})(2-e^{a(\tau_+-\tau_-)}(9-e^{a(\tau_+-\tau_-)}))\over 
(1+e^{a(\tau_+-\tau_-)})^3}
\biggr],
\nonumber\\
\label{expressionFq}
\end{eqnarray}
where we defined $\varepsilon=|\tau_-^x -\tau_-^y|$, and 
$P$ and $a(\tau_+-\tau_-)$ are functions of $q$ related by 
$P(q)=q/\sqrt{1+q^2}=-\tanh a(\tau_+-\tau_-)/2$.
Note also that $q$ is a function of the coordinates written as Eq.~(\ref{qtt}), or 
$q = a(t-r - 1/2a^2r)/\sin{\theta}$. In $F(q)$ there are two terms that diverge in 
the coincidence limit $\varepsilon=0$. This divergence is due to the short-distance 
motion of the particle, which originates from our formulation based on the point 
particle (see also Ref.~\cite{OYZ15}). The divergence due to the short-distance 
motion of the particle can be removed by taking the finite size effect of the particle 
into account. For simplicity, we omit the divergent terms; however, this prescription 
does not alter our conclusions as long as the cutoff value is $a\varepsilon={\cal O}(1)$. 
Figure 2 plots $G(q)$ and $F(q)$ as functions of $q$. 
With the diverging terms removed, the classical part $f^C$ reduces to 
the classical Larmor radiation, while the quantum part $f^Q$  goes to the quantum radiation. 
As in the case of the massless scalar field, the 
quantum part $f^Q$ is smaller than the classical counterpart $f^C$ by one order in $a/m$. 
However, the angular distribution for the electromagnetic field case is quite different 
from that for the case of the massless scalar field, as described below. 

\begin{figure}[t]
  \begin{center}
   \includegraphics[width=67mm]{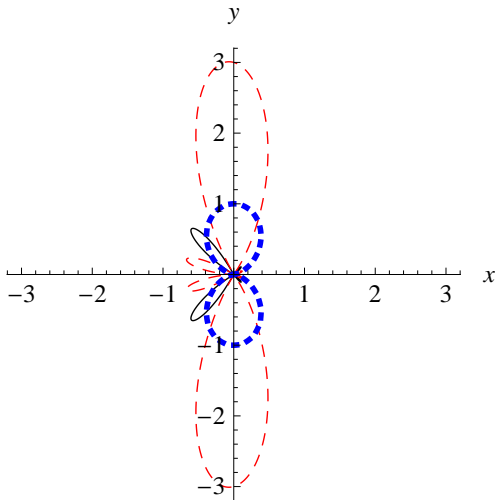}
  \end{center}
   \vspace{-0.5cm}
 \caption{
Angular distribution of the classical radiation $\sin^{-4}{\theta} G(\tau_-^x ,\theta)$ (blue dotted curve) and the quantum radiation $\sin^{-4}{\theta} F(\tau_-^x ,\theta)$ (black solid: positive values; red dashed curve: negative values) at $a \tau_-^x = 0$. The coordinates $x$ and $y$ are $x^1$ and $\sqrt{(x^2)^2+(x^3)^2}$, respectively. 
\label{fig:sixx}}
\end{figure}

Figure 3 shows the polar plot of 
$\sin^{-4}{\theta}G(q(\tau_-^x, \theta))$ (blue dotted curve) and 
$\sin^{-4}{\theta}F(q(\tau_-^x, \theta))$ (black solid curve for {\em positive} values and red dashed curve for {\em negative} values) with $\tau_-^x$ fixed as $a \tau_-^x=0$. This plot is made by regarding $q$ as a function of $\tau_-^x$ and $\theta$, i.e.,
\begin{eqnarray}
q(\tau_-^x,\theta) = \sinh{\left[ a \tau_-^x - \text{arctanh}{\left( \cos{\theta} 
\right)} \right]}.
\label{qtt}
\end{eqnarray}
Fig.~3 describes the angular distribution of the energy flux at the moment $a\tau_-^x=0$.
It is well known that the classical energy flux  $f^C$ of the Larmor radiation is dominantly emitted 
perpendicular to the direction of acceleration. The quantum radiation flux is almost 
entirely negative, although some small positive regions exist. The emission directions in the dominant 
regions are similar to those of the classical radiation. This is understood as the suppression of the 
Larmor radiation due to the quantum effect, which is consistent with the predictions of the model based 
on a particle and a massless scalar field \cite{OYZ15}. It is also consistent with studies on the 
quantum correction to the Larmor radiation \cite{HW,NSY,YN,KNY,NY},  though our approach described 
here is quite different from those studies.
The interference effect plays an important role for this property because the quantum
radiation comes from the interference terms; however, this also makes it difficult 
to understand the results in an intuitive manner.
Negative quantum energy density may appear in the quantum field theory, 
depending on quantum state \cite{Ford:2009vz,Ford:2007}. 
Following one such possible explanation, the quantum radiation may contain quantum-correlated photons
distinct from the classical radiation. 
This is an interesting possibility, but it is out of the scope of the present paper. 
However, it should be stressed that the total radiation combining the classical Larmor 
radiation and the quantum radiation is positive unless the condition $a/m\ll1$ is broken. 

\section{Summary and Conclusions}
We have investigate the properties of quantum radiation produced by a uniformly accelerating charged 
particle undergoing thermal random motions, which originates from the coupling to the 
vacuum fluctuations of the electromagnetic field. The thermal random motions are 
regarded to result from the Unruh effect.
The energy flux of the quantum radiation is negative and smaller than that of Larmor radiation by one order 
in $a/m$,  where $a$ is the constant acceleration  and $m$ is the mass of the particle. 
Thus, the quantum radiation appears to be a suppression of the classical Larmor radiation. 
These properties of the quantum radiation might be interesting because 
it may be possible to 
experimentally verify them. In Ref.~\cite{IYZ}, possible methods of experimentally testing 
the Unruh effect are discussed.
One difficulty comes from the thermalization time, $2/a\sigma=12\pi m/(a^2e^2)$, 
which is quite long; additionally, it is difficult to 
keep an electron in a state of acceleration for a sufficient duration of time~\cite{IYZ}. 
It is interesting and important to investigate phenomena during relaxation process.
One will find corrections by considering a particle accelerated in finite duration of time but 
not eternal. This subject will be analyzed in a future work.
However, the assumption of the eternal acceleration will not drastically change our results 
when the acceleration time is much longer than the relaxation time. 
This is confirmed at least for the random motions of a charged particle~\cite{IYZ}.

There are other problems that need to be investigated further, the first of which is the 
possible contamination of the longitudinal fluctuations. 
In the present paper, we have only investigated the transverse random motions of the 
particle, which follow  the energy 
equipartition relation. Longitudinal fluctuations are motions in the $x^1$ direction. 
Longitudinal random motions do not 
follow  the energy equipartition relation, and the variance of the velocity is of 
the order $(a/m)^3$, as in the case of 
the scalar field~\cite{IYZ,OYZ14}. Therefore, the longitudinal mode may not make a 
significant contribution to the quantum radiation. 

Another problem might come from the divergent terms in the energy-momentum tensor 
and the energy flux, which appear due to our theoretical framework based on the 
point particle, reflecting its short-distance dynamics \cite{OYZ14}. 
Here, we have assumed that the divergent terms can be 
removed by taking into account the finite size effect. 
More careful discussion about this divergent term might be necessary. 
However, we can observe the following point. One can read that the divergent 
terms are odd functions of $P$ (see Eq.~(\ref{expressionFq})). This property is the same as that of the model 
consisting of a particle and a scalar field in Ref.~\cite{OYZ15}. 
This means that the divergent terms contribute to the energy flux as odd functions 
of $t-r$ at a large distance, which vanish if one integrates them over the time. 

Finally, it is useful to compare our results with those of the previous
works \cite{ChenTajima,Schutzhold,Schutzhold2}, 
which investigated the Unruh radiation from an accelerated charged particle. 
As noted in the first part of this paper, the different point of the present 
paper is the inclusion of the interference term. 
In the two point function of electromagnetic field, the term 
$\langle A_{\rm inh}^\alpha(x)A_{\rm inh}^\beta(y)\rangle$ 
cancels out. Therefore, the non-trivial signature of our results comes from the 
remaining interference term. 
This has resulted in the difference between our results and the previous works. 
For our reference, we have investigated the flux which comes from the term 
$\langle A_{\rm inh}^\alpha(x)A_{\rm inh}^\beta(y)\rangle$
(see the appendix for details). The energy flux from this term is positive, 
and is dominantly emitted in the direction of the acceleration.
These features do not contradict with those reported in Refs.
\cite{ChenTajima,Schutzhold,Schutzhold2}, which are quite opposite 
to those of Eq.~(\ref{fluxc}).

\begin{acknowledgments}
We would like to thank J. Yokoyama, T. Suyama, K. Fukushima, and G. Matsas 
for helpful  discussions and comments. We also thank 
Prof. P. Chen for critical discussions that took place at the beginning of this study. 
N.O. is supported by a research program of the Advanced Leading Graduate Course for Photon Science 
(ALPS) at the University of Tokyo.
The research by K.Y. is supported by a Grant-in-Aid for Scientific Research of Japan Ministry of 
Education, Culture, Sports, Science and Technology (No.15H05895). 
\end{acknowledgments}

\appendix
\section{Screened inhomogeneous term}
\begin{figure}[b]
  \begin{center}
   \includegraphics[width=67mm]{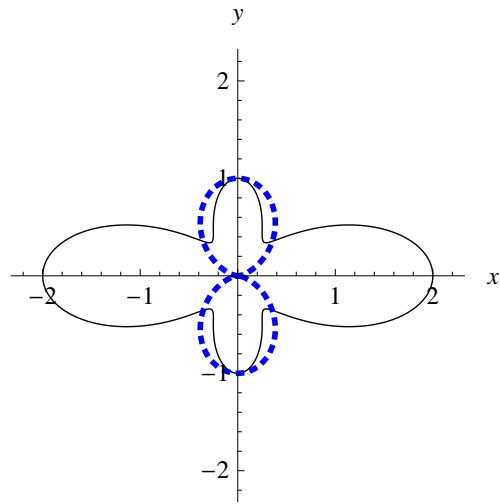}
  \end{center}
   \vspace{-0.5cm}
 \caption{Same as figure \ref{fig:sixx} but for $\sin^{-4} \theta H(\tau_-^x,\theta)$ (solid line) and 
$\sin^{-4} \theta G(\tau_-^x,\theta)$ (dashed line). 
\label{inhinhf}}
\end{figure}

We find that the two point function $\langle A_{\text{inh}}^\alpha(x) A_{\text{inh}}^\beta(y) \rangle$
is explicitly written as 
\begin{eqnarray}
&&\langle A_{\text{inh}}^\alpha(x) A_{\text{inh}}^\beta(y) \rangle
\nonumber\\
&&~~~~~~
=-{e^2\eta_{ij}E^{\alpha i}_{(-)}(x)E^{\beta j}_{(-)}(y)\over (4\pi)^2\rho_0(x)\rho_0(y)}
{a\over 2\pi m},
\label{infinfinf}
\end{eqnarray}
which cancels due to the interference term in our computation. 
However, it might be useful to compute the flux from this term. 
We find the following flux from Eq.~(\ref{infinfinf}),
\begin{eqnarray}
f^S = \left( \frac{e}{4 \pi} \right)^2 \frac{a^3}{2 \pi m} \frac{1}{r^2} \frac{1}{\sin^4{\theta}} H(q)
\end{eqnarray}
with $H(q)$ defined by 
\begin{eqnarray}
H(q) = \frac{1}{(1+q^2)^2} \left( 2P^2 - \frac{4 P^2}{1+q^2} + \frac{1}{1 +q^2} \right).
\end{eqnarray}
Figure \ref{inhinhf} compares the angular distribution of the classical radiation 
$\sin^{-4}{\theta} G(\tau_-^x ,\theta)$ (blue dotted curve) and the 
quantum radiation $\sin^{-4}{\theta} H(\tau_-^x ,\theta)$ 
(black solid: positive values) at $a \tau_-^x = 0$.
The screened radiation $f^S$ is dominantly emitted in the direction of the acceleration. 


\end{document}